\documentclass[preprint,12pt]{elsarticle}

\usepackage{graphics}
\usepackage{graphicx}
\usepackage{epsfig}
\usepackage{amssymb}

\journal{Applied Radiation and Isotopes}

\begin{document}

\begin{frontmatter}

\title{Half-life measurement of $^{65}$Ga with $\gamma$-spectroscopy}

\author[ATOMKI]{Gy. Gy\"urky\corref{cor}}
\ead{gyurky@atomki.mta.hu}
\author[ATOMKI]{Z. Hal\'asz}
\author[ATOMKI]{G.G. Kiss}
\author[ATOMKI]{T. Sz\"ucs}
\author[ATOMKI]{Zs. F\"ul\"op}
\address[ATOMKI]{Institute of Nuclear Research (Atomki), H-4001 Debrecen, POB.51., Hungary}
\cortext[cor]{corresponding author}
\begin{abstract}

The literature half-life value of $^{65}$Ga is based on only one experiment carried out more than 60 years ago and it has a relatively large uncertainty. In the present work this half-life is determined based on the counting of the $\gamma$-rays following the $\beta$-decay of $^{65}$Ga. Our new recommended half-life is t$_{1/2}$\,=\,(15.133\,$\pm$\,0.028)\,min which is in agreement with the literature value but almost one order of magnitude more precise.

\end{abstract}


\end{frontmatter}

\section{Introduction}
\label{sec:introduction}

The half-lives of most of the relatively long lived radioactive isotopes close to the valley of $\beta$-stability were typically determined back in the 1950s and 1960s. Due in part to the immature experimental technique of that time, the precision of those old measurements is often insufficient and values are sometimes found which turn out to be incorrect \cite[for example]{far11}. In the case of isotopes which do not have importance for a specific application (e.g. isotopes for medical purposes), it may happen that the literature value is based on only one, typically rather old experiment. $^{65}$Ga, studied in the present work, is one of such isotopes. Its adopted half life, t$_{1/2}$\,=\,(15.2\,$\pm$\,0.2)\,min \cite{NDS111}, is based on one experimental result published in 1957 \cite{Dan57} in German language providing limited experimental and data analysis information.

Recently, the cross section of the $^{64}$Zn(p,$\gamma$)$^{65}$Ga reaction was measured with the activation method, i.e. the number of reactions was determined based on the decay measurement of the produced isotope \cite{gyu14}. On the one hand, the precision and reliability of such an experiment is directly related to the knowledge and accuracy of the half-life value of the reaction product. On the other hand, such an activation experiment provides the possibility of a new determination of the half-life.

This paper reports on the half-life measurement of $^{65}$Ga which was carried out as a parallel project of the $^{64}$Zn(p,$\gamma$)$^{65}$Ga cross section measurement. In Section \ref{sec:experimental} the experimental technique is described while Section \ref{sec:results} contains the details of the data analysis and the obtained result. The new recommended half-life value is given in the summary Section \ref{sec:summary}.

\section{Experimental procedure}
\label{sec:experimental}

\subsection{Source preparation}

The $^{65}$Ga sources were prepared with the $^{64}$Zn(p,$\gamma$)$^{65}$Ga reaction using the cyclotron accelerator of Atomki. $^{64}$Zn target were prepared by evaporating isotopically enriched (99.71\,\%) $^{64}$Zn onto thin (2\,$\mu$m) Al foils. The targets were irradiated by proton beams of two different energies (4.5 and 6.0\,MeV) with typical beam intensities of 2\,$\mu$A. The irradiations were roughly one hour long (about four half-lives of $^{65}$Ga). 

Altogether eight sources were produced for the half-life measurements, four at both beam energies. Besides $^{65}$Ga, the only radioisotope observed in the targets in measurable quantity was $^{61}$Cu (t$_{1/2}$\,=\,3.339\,$\pm$\,0.008\,h \cite{NDS125}) produced by the $^{64}$Zn(p,$\alpha$)$^{61}$Cu reaction. The measurement of this isotope was used to assess the stability of the detection system and to estimate the systematic uncertainties (see Section \ref{sec:results}). After the irradiations the targets were transported to the counting facility where the decay measurements started typically 15 minutes after the end of the activations.

\subsection{Gamma-detection}

$^{65}$Ga decays by positron emission or electron capture and the decay populates a large number of excited states in $^{65}$Zn. The decay is thus followed by the emission of many $\gamma$-rays belonging to various transitions. The levels involved in the decay are all short-lived (on the $\mu$s level or below) compared to the $\beta$-decay lifetime of $^{65}$Ga, therefore by measuring the $\gamma$-radiation the $^{65}$Ga half-life can be determined.

The five most intense $\gamma$-transitions were selected and analyzed for the half-life determination. These transitions are listed in Table\,\ref{tab:transitions} where the energies as well as the relative intensities of the transitions are shown. The table also shows the two measured transitions of the $^{61}$Cu decay.

\begin{table}
\centering
\caption{\label{tab:transitions} Energies and relative intensities of the transitions used for the analysis}
\begin{tabular}{lll}
\hline
Isotope & E$_\gamma$/keV & Rel. int. [\%] \\
\hline
$^{65}$Ga & 54  &	4.9 \\
& 61 &	11.4 \\
& 115 &	54 \\
& 153 &	8.9 \\
& 752  &	8.1 \\
\hline
$^{61}$Cu & 283 & 12.2 \\
& 656 & 10.8 \\
\hline
\end{tabular}
\end{table}

The $\gamma$-spectra were measured with two Canberra HPGe detectors, both having 100\,\% relative efficiency. For runs \#1\,--\,\#4 the first detector with U-style cryostat configuration \cite{canberra} and complete 4\,$\pi$ low background, multilayer shielding was used. The second detector used for runs \#5\,--\,\#8 had a horizontal dipstick cryostat configuration and the setup was shielded with 5\,cm lead. 

The spectra were collected with ORTEC Model ASPEC-927 multichannel analyzers using the ORTEC MAESTRO software \cite{ortec}. One minute spectra were stored and the decay of the sources was followed for typically 2 hours. The maximum dead time of the counting systems at the beginning of the measurements was below 10\,\% which decreased to a negligible level during the runs.

\begin{figure}
\centering
\resizebox{0.7\textwidth}{!}{\rotatebox{270}{\includegraphics{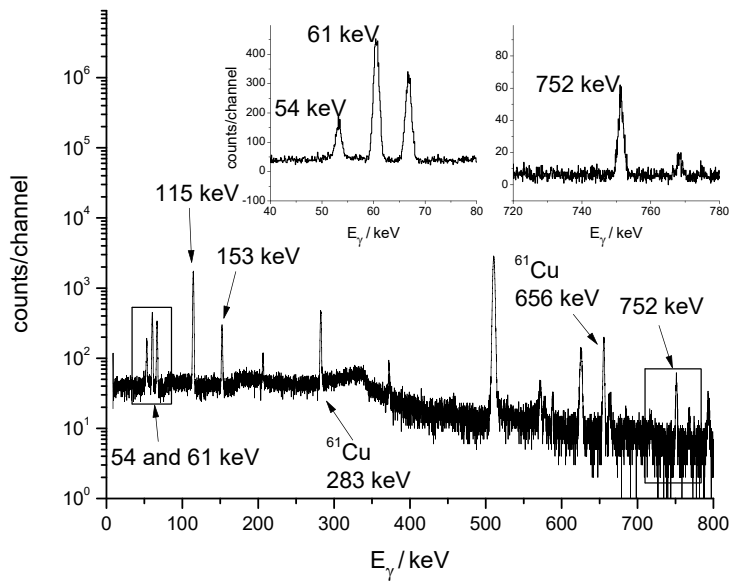}}}
\caption{\label{fig:spectrum} Gamma spectrum measured in run \#1 for one minute. The peaks used for the analysis are indicated. The insets show two relevant parts of the spectrum in linear scale.}
\end{figure}

Fig.\,\ref{fig:spectrum} shows a typical $\gamma$-spectrum measured at the the beginning of run \#1 for one minute. The peaks used for the analysis are indicated. The insets show two relevant parts of the spectrum in linear scale. All used peaks are well separated from the other peaks. The peak areas could therefore be determined by peak integration fixing a linear background using two peak-free regions directly on each sides of the studied peaks. 

\section{Data analysis and results}
\label{sec:results}

The determined peak areas were corrected for the dead time of the counting system as given by the MAESTRO software. The dead time determination of MAESTRO was found to be reliable in previous works \cite{gyu09,gyu12} but has also been checked in the present work (see below). The dead time corrected peak area as a function of time was fitted with an exponential function following the procedure described in \cite{leo94}.

\begin{figure}
\centering
\resizebox{0.7\textwidth}{!}{\rotatebox{270}{\includegraphics{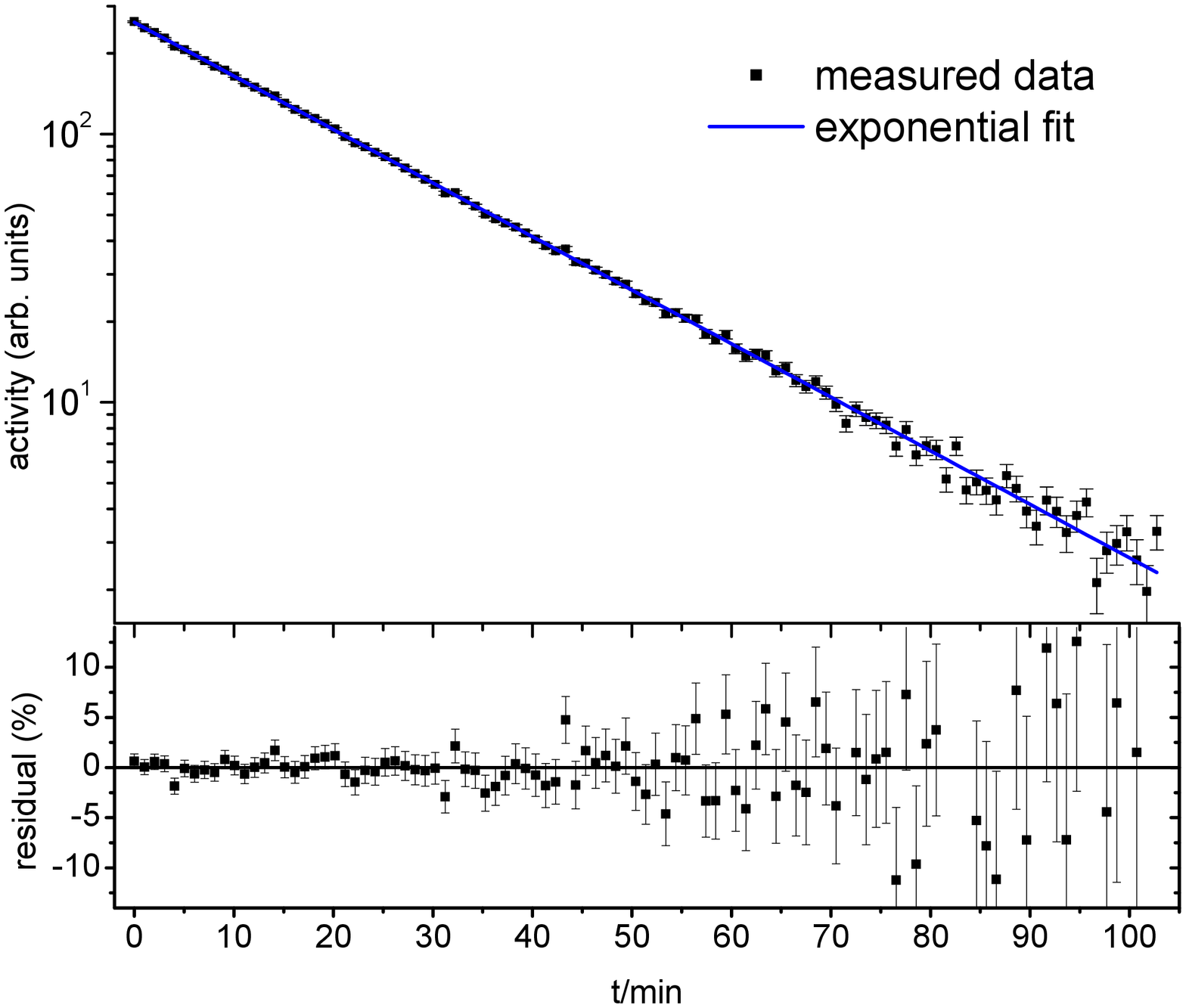}}}
\caption{\label{fig:decay} Measured and fitted source activity in run \#5 based on the 115\,keV transition.}
\end{figure}

A typical decay curve and the fit residuals are shown in Fig.\,\ref{fig:decay}. As it can be seen qualitatively from the lack of structure in the residual values, no deviation from the exponential decay was observed. This will be further analyzed quantitatively below. Taking into account the eight runs and the five analyzed transitions, 40 statistically independent half-life values were obtained. These are listed in Table\,\ref{tab:results} where the uncertainties are statistical only stemming from the fit parameter determination. The reduced chi square ($\chi^2_{\rm red}$) values of the fits are also indicated, the values scatter around one proving the goodness of the exponential fit. The weighted average of the 40 runs gives t$_{1/2}$\,=\,(15.133\,$\pm$\,0.009)\,min. The $\chi^2_{\rm red}$ of the averaging is 3.10,  higher than one, therefore the statistical uncertainty of the averaged half-life has been increased by $\sqrt{\chi^2_{\rm red}}$ resulting in t$_{1/2}$\,=\,(15.133\,$\pm$\,0.016)\,min. Note, that this uncertainty is the same as if the external variance of the data points had been calculated as described for example in \cite{gilmore}.

\begin{table}
\centering
\footnotesize
\caption{\label{tab:results} Measured half-lives. The given uncertainties are statistical only. See text for details.}
\begin{tabular}{lrr@{\,}c@{\,}lc}
\hline
run no. & transition & \multicolumn{3}{c}{t$_{1/2}$/min} & $\chi^2_{\mathrm{red}}$ \\
\hline
\#1	&	54 keV	&	15.795	&	$\pm$	&	0.317	&	1.04	\\
	&	61 keV	&	15.112	&	$\pm$	&	0.100	&	1.06	\\
	&	115 keV	&	15.137	&	$\pm$	&	0.029	&	0.98	\\
	&	153 keV	&	14.619	&	$\pm$	&	0.140	&	0.88	\\
	&	752 keV	&	15.716	&	$\pm$	&	0.216	&	1.57	\\
	\hline
\#2	&	54 keV	&	15.085	&	$\pm$	&	0.482	&	1.11	\\
	&	61 keV	&	15.266	&	$\pm$	&	0.172	&	0.96	\\
	&	115 keV	&	15.203	&	$\pm$	&	0.056	&	0.84	\\
	&	153 keV	&	15.057	&	$\pm$	&	0.269	&	1.14	\\
	&	752 keV	&	15.592	&	$\pm$	&	0.326	&	0.98	\\
	\hline
\#3	&	54 keV	&	15.407	&	$\pm$	&	0.229	&	1.23	\\
	&	61 keV	&	15.148	&	$\pm$	&	0.086	&	0.82	\\
	&	115 keV	&	14.921	&	$\pm$	&	0.037	&	1.21	\\
	&	153 keV	&	15.274	&	$\pm$	&	0.112	&	1.11	\\
	&	752 keV	&	15.124	&	$\pm$	&	0.227	&	1.01	\\
	\hline
\#4	&	54 keV	&	15.141	&	$\pm$	&	0.181	&	1.10	\\
	&	61 keV	&	15.091	&	$\pm$	&	0.066	&	1.06	\\
	&	115 keV	&	15.193	&	$\pm$	&	0.029	&	1.21	\\
	&	153 keV	&	15.148	&	$\pm$	&	0.092	&	0.75	\\
	&	752 keV	&	15.235	&	$\pm$	&	0.164	&	1.26	\\
	\hline
\#5	&	54 keV	&	15.740	&	$\pm$	&	0.363	&	0.86	\\
	&	61 keV	&	15.151	&	$\pm$	&	0.122	&	0.77	\\
	&	115 keV	&	15.076	&	$\pm$	&	0.034	&	0.89	\\
	&	153 keV	&	14.952	&	$\pm$	&	0.153	&	1.01	\\
	&	752 keV	&	15.300	&	$\pm$	&	0.209	&	1.29	\\
	\hline
\#6	&	54 keV	&	16.122	&	$\pm$	&	0.764	&	0.99	\\
	&	61 keV	&	15.574	&	$\pm$	&	0.255	&	1.37	\\
	&	115 keV	&	14.785	&	$\pm$	&	0.075	&	1.06	\\
	&	153 keV	&	16.310	&	$\pm$	&	0.449	&	1.28	\\
	&	752 keV	&	14.761	&	$\pm$	&	0.377	&	0.88	\\
	\hline
\#7	&	54 keV	&	15.050	&	$\pm$	&	0.190	&	1.17	\\
	&	61 keV	&	15.176	&	$\pm$	&	0.070	&	1.01	\\
	&	115 keV	&	15.146	&	$\pm$	&	0.028	&	1.12	\\
	&	153 keV	&	15.139	&	$\pm$	&	0.092	&	0.99	\\
	&	752 keV	&	14.852	&	$\pm$	&	0.146	&	0.78	\\
	\hline
\#8	&	54 keV	&	15.267	&	$\pm$	&	0.103	&	1.03	\\
	&	61 keV	&	15.167	&	$\pm$	&	0.042	&	1.12	\\
	&	115 keV	&	15.152	&	$\pm$	&	0.017	&	1.08	\\
	&	153 keV	&	15.188	&	$\pm$	&	0.056	&	1.00	\\
	&	752 keV	&	15.198	&	$\pm$	&	0.102	&	0.93	\\
	\hline
\hline
\multicolumn{2}{l}{error weighted mean} & 15.133 &	$\pm$ & 0.009 & 3.10 \\
\hline
\end{tabular}
\end{table}

\begin{figure}
\centering
\resizebox{0.7\textwidth}{!}{\rotatebox{270}{\includegraphics{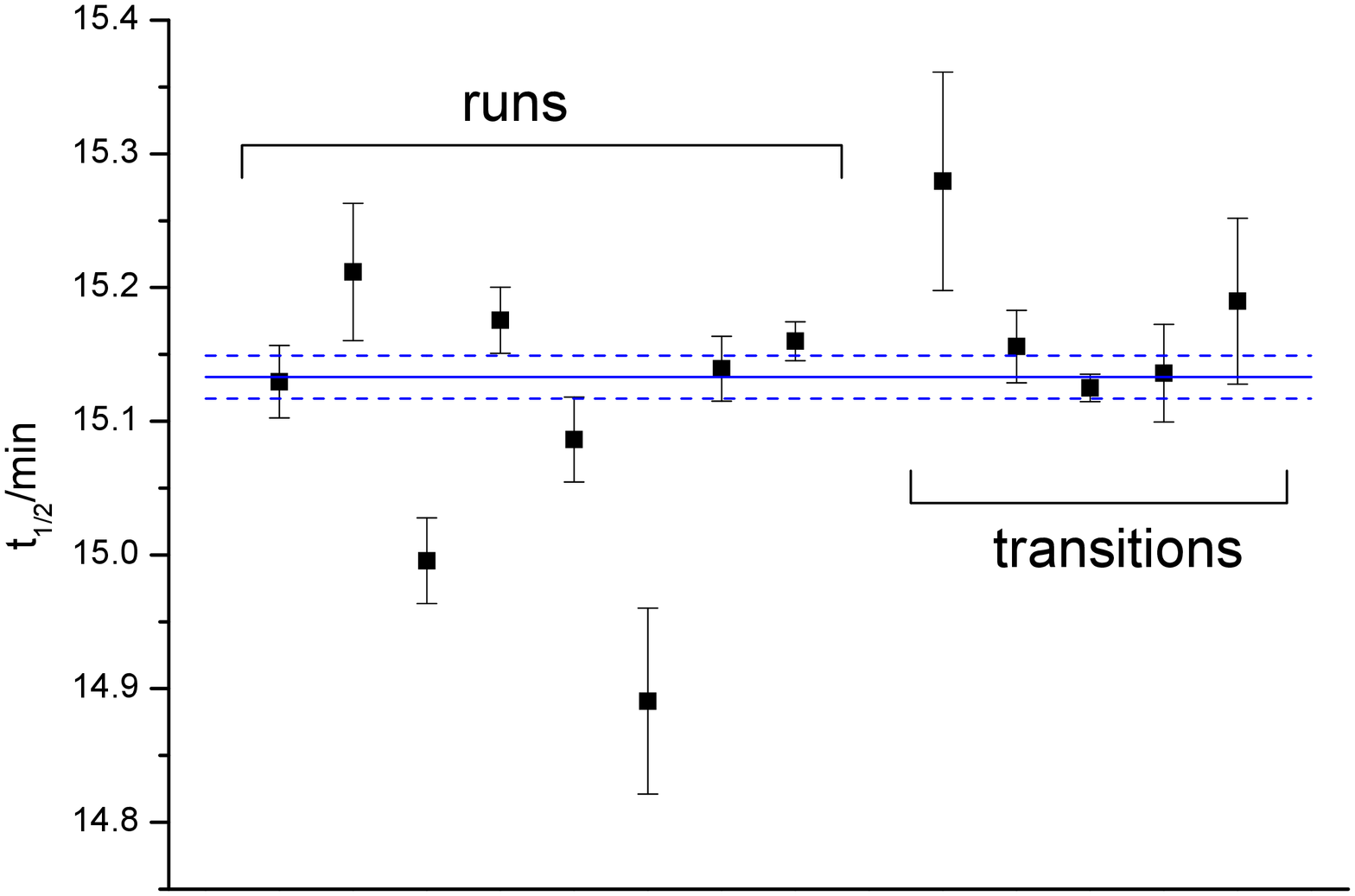}}}
\caption{\label{fig:results} (Left) averaged half-life results of the runs from \#1 left to \#8 right (averaged over the 5 measured transitions) and (right) of the five transitions (averaged over the eight runs). The plotted error bars are statistical only. The final half-life value and its statistical uncertainty is also show as horizontal lines.}
\end{figure}

Besides the statistical uncertainty, the determined half-life value may be prone to systematic uncertainties. This possibility is studied by the following. It was checked whether the determined half-life shows any systematic difference depending on the detection system used or on the $\gamma$-transition measured. This test is represented in Fig.\,\ref{fig:results} where the results averaged over the $\gamma$-transitions in the various runs (left side) and over the runs for a given transition (right side) are plotted. In the case of the different transitions the final result is dominated by the strongest one (115 keV) which always has the highest statistics, but the other transitions result in half-life values in agreement with the average (four values are within one, and one value is within two standard deviations from the average). As for the different runs, the statistical uncertainties are again somewhat different owing to the different initial activities of the sources. The results obtained with the two detectors\footnote{The average half-life values for the first and second detectors are (15.124\,$\pm$\,0.015)\,min and (15.138\,$\pm$\,0.011)\,min, respectively. Here the uncertainty values are not increased by $\sqrt{\chi^2_{\rm red}}$.}, however, are in agreement within the statistical uncertainty.

Since the data acquisition system including the dead time determination was the same for all runs and transitions, a possible systematic uncertainty related to the acquisition system must be assessed. For this purpose the measurement of the $^{61}$Cu decay was used. This isotope has a precisely known half-life of (3.339\,$\pm$\,0.008)\,h \cite{NDS125}. In the case of the sources that were prepared with the 6\,MeV proton beam (runs \#1, \#2, \#5 and \#6), the activity of $^{61}$Cu was high enough for a precise half-life analysis. The results are shown in Table\,\ref{tab:61Cu} for the four runs and the two measured transitions of 283 and 656\,keV. The average of the results is in excellent agreement with the precise literature value.

\begin{table}
\centering
\caption{\label{tab:61Cu} Half-life of $^{61}$Cu determined for the estimation of systematic uncertainties. The given uncertainties are statistical only.}
\begin{tabular}{lrr@{\,}c@{\,}l}
\hline
run no. & transition & \multicolumn{3}{c}{t$_{1/2}$/h}  \\
\hline
\#1	&	283 keV	&	3.385	&	$\pm$	&	0.041	\\
\#2	&	283 keV	&	3.261	&	$\pm$	&	0.128	\\
\#5	&	283 keV	&	3.535	&	$\pm$	&	0.079	\\
\#6	&	283 keV	&	3.363	&	$\pm$	&	0.129	\\
\#1	&	656 keV	&	3.284	&	$\pm$	&	0.048	\\
\#2	&	656 keV	&	3.258	&	$\pm$	&	0.151	\\
\#5	&	656 keV	&	3.090	&	$\pm$	&	0.073	\\
\#6	&	656 keV	&	3.141	&	$\pm$	&	0.139	\\
	\hline
\multicolumn{2}{l}{error weighted mean} &	3.321	&	$\pm$	&	0.025	\\
	\hline
\multicolumn{2}{l}{literature}	&	3.339	&	$\pm$	&	0.008	\\
\hline
\end{tabular}
\end{table}

As a conservative estimate of the systematic uncertainty, the following procedure was adopted. The dead time indicated by the acquisition system was artificially modified until the average measured $^{61}$Cu half-life changed as much as its statistical uncertainty (0.025\,h, see Table\,\ref{tab:61Cu}). For such a change, 16\,\% relative modification of the dead time value was needed. Then this dead time modification was applied to all the individual spectra used for the half-life determination of $^{65}$Ga (the determined peak areas were corrected using the modified dead times) and its effect on the half-life value was checked. The resulting half-life differed from the original one by 0.023\,min. This value is taken as the systematic uncertainty of the determined $^{65}$Ga half-life.

\section{Summary and the recommended $^{65}$Ga half-life}
\label{sec:summary}

Based on the average of 40 statistically independent measurement, the half-life of $^{65}$Ga was found to be 15.133\,min with a statistical uncertainty of 0.016\,min. From the measurement of $^{61}$Cu decay, a systematic uncertainty of 0.023\,min is obtained for $^{65}$Ga. Combining quadratically the two uncertainties, our new half-life value is t$_{1/2}$\,=\,(15.133\,$\pm$\,0.028)\,min. This result is in agreement with the literature value but roughly a factor of 7 more precise. This value changes only marginally if averaged with the literature value. We recommend therefore to use our new result as the adopted half-life of $^{65}$Ga.

\section*{Acknowledgements}
 
This work was supported by NKFIH grants K120666 and NN128072 and by the \'UNKP-18-4-DE-449 New National Excellence Program of the Human Capacities of Hungary. G.G. Kiss acknowledges support form the J\'anos Bolyai research fellowship of the Hungarian Academy of Sciences.


\end{document}